\documentclass[12pt]{spieman} 
\usepackage[]{graphicx}
\usepackage{setspace}
\usepackage{tocloft}

\usepackage{lineno}

\usepackage{caption}
\usepackage{subcaption}
\usepackage{amsmath}
\usepackage{array,multirow}
\usepackage{xcolor}

\linenumbers

\title{Astrometric Errors Introduced by Interpixel Capacitive Coupling in Hybridized Arrays} 

\author{Kevan Donlon\supscr{a}, Zoran Ninkov\supscr{a}, Stefi Baum\supscr{a,b}}

\affiliation{\supscrsm{a}Rochester Institute of Technology, Chester F. Carlson Center for Imaging Science, 54 Lomb Memorial Drive, Rochester, New York, 14623\\
\supscrsm{b}University of Manitoba, Department of Physics and Astronomy, 66 Chancellors Cir, Winnipeg, MB, Canada}

\cftpagenumbersoff{figure}
\cftpagenumbersoff{table} 
\begin{document} 
\maketitle 

\begin{abstract}
Interpixel capacitance (IPC) between adjacent pixels in hybridized arrays gives rise to an electrostatic cross talk. This cross talk causes MTF degradation and blurring of images or spectra collected using these devices. As pixel size is driven down from the 18 micron pixel pitch of the H2RG read out circuits to the 10 or 15 micron H4RGs IPC is driven up resulting in greater cross talk, all else being equal. Mounting evidence indicates that IPC varies as a function of depletion state of the photo-active diodes. For single pixel events, increasing the event intensity corresponds to a decreasing fractional coupling.  If left uncorrected, IPC can give rise to systematic errors in precision astrometric and photometric measurements, in particular when dealing with confused point sources or spatially extended structures for shape measurements as demonstrated through comparison of registered sources from ESO HAWK-I and HST ACS WFC datasets. Furthermore these errors will be the most significant when operating near the sensitivity limit of these devices. Deconvolution based correction methods are invalidated by this same signal dependence. Instead a numerical method of successive approximation can be used to correct coupling due to a well characterized IPC. Examination of single pixel reset data above flat fields could be used to characterize IPC's functional relationship for neighboring pixels.  This higher quality characterization can result in more accurate correction.
\end{abstract}


{\noindent \footnotesize{\bf Address all correspondence to}: Kevan Donlon, Rochester Institute of Technology, Chester F. Carlson Center for Imaging Science, 54 Lomb Memorial Drive, Rochester, New York, 14623; E-mail: \linkable{KevanADonlon@gmail.com} }

\begin{spacing}{2} 

\label{sect:intro}
	Hybridized detector arrays are composed of three distinct regions:  a photo-sensitive layer,  a read-out circuit (ROIC) layer, and an indium bump bond layer responsible for connecting the two.  This detector design divorces the read-out electronics from the light detection which allows for unconventional materials or detection techniques to be used without disrupting the proven technology in the ROIC~\cite{Rogalski00}.

	The introduction of an intermediate layer that supports the presence of electric fields gives rise to a conventional capacitor; conductive materials separated by a dielectric epoxy.  This parasitic electrostatic connection causes the electronic state of a single pixel to impact not just the readout node that corresponds to that pixel, but also the read-out node corresponding to nearby pixels~\cite{Moore06}$^,$\cite{Ohanian07}.  In addition to this conventional capacitor, electric fields from the depletion region of the photodiodes bleed out into the bump bond layer\cite{Moore06}$^,$\cite{Donlon17}.  These fringing fields vary non-linearly as a diode is exposed to photons\cite{Cheng09}$^,$\cite{Donlon16}$^,$\cite{Donlon17}.  This non-linear field strength results in a non-uniform capacitance between neighboring pixels as their depletion region changes size.

	The impact that this capacitance has on an output image, $O(i,j)$, has historically been formalized as convolution of an input image, $I(i,j)$, with a nearest neighbor coupling kernel, $K(i,j)$\cite{Moore04}.  The coupling coefficient, $\alpha$, indicates the fractional signal collected on a single pixel that manifests on the output of each neighboring pixel.
\begin{equation}
\label{eq:blur}
K(i,j)=\left(\begin{array}{ccc} 0&\alpha&0\\\alpha&1-4\alpha&\alpha\\0&\alpha&0\end{array}\right)
\end{equation}
\begin{equation}
\label{eq:conv}
O(i,j) = K(i,j) \ast I(i,j)
\end{equation}
	A non-uniform capacitance results in a non-uniform coupling coefficient.  Due to the fringing fields from the photo-diodes, IPC varies as a function of photons collected in the neighborhood of readout node~\cite{Donlon17}.  Only giving consideration to the final image output, coupling between two pixels varies as a function of the level of each pixel.  This gives rise to a systems of equations where $S$ is the uncoupled incident level of the central pixel,  $B$ is the background signal, $N$ is the readout value of the neighboring pixel, and $C$ is the readout value of the central pixel and $\alpha$ is the fractional coupling coefficient.
\begin{equation}
\label{eq:algebra_system_1}
C = S - 4 \alpha S + 4 \alpha B
\end{equation}
\begin{equation}
\label{eq:algebra_system_2}
N = B + \alpha S -\alpha B
\end{equation}
Solved for $\alpha$ this yields:
\begin{equation}
\label{eq:algebra_solved}
\alpha = \frac{N-B}{C+4N-5B}
\end{equation}
 Using this convention, $\alpha = \alpha(S,B)$.
	Within this work, a method to characterize this coupling is discussed, a technique for correcting a well characterized coupling is referenced and applied to data for the scientific ramifications to be assessed. The domain of these parameters is intentionally left ambigious.  All values can be taken in the charge domain, the voltage domain, or any arbitrarily defined continious injective mapping over the charge domain.  The key facet, is that the domain must be consistent and must match the domain for the coupling coefficient.  The particular functional behavior of $\alpha$ will be distinct if given in terms of $Q$ or $V$, but the preceding equations and the conclusions that follow will hold in any particular instance of this general approach.

\section{IPC and systematic errors}
\label{sect:star_distortions}
Characterization of the coupling in isolation can aid in understanding the nature of the measurement tool and the systematic errors that it can introduce, but perhaps more valuably when that error behaves deterministically rather than stochastically, its impact on scientific data in particular circumstances can be elucidated and corrected.  Previous work has made an effort to predict the impact that IPC can have on scientific data when examining crowded fields~\cite{Donlon18};  here those predictions will be empirically verified.
With a well characterized coupling coefficient, the effect of IPC can be backed out using the following iterative method~\cite{Donlon18} with $S$, $C$, and $\alpha$ as used in equations \ref{eq:algebra_system_1} to \ref{eq:algebra_solved}, $q$ indicating iteration count, and $m$, $n$, $i$, and $j$ indicating spatial pixel indexes:

\begin{equation}
\begin{split}
\label{eq:full_dec}
\hat{S}_{q}(i,j) = &C(i,j) - \sum_{m=i-1}^{i+1} \sum_{n=j-1}^{j+1} \biggl[ \alpha\biggl(\hat{S}_{q-1}(i,j),\hat{S}_{q-1}(m,n)\biggr) \cdot \hat{S}_{q-1}(m,n)\biggr]  \\& + \sum_{m=i-1}^{i+1} \sum_{n=j-1}^{j+1}\biggl[\alpha\biggl(\hat{S}_{q-1}(m,n),\hat{S}_{q-1}(i,j)\biggr) \cdot \hat{S}_{q-1}(i,j)\biggr]
\end{split}
\end{equation}
\begin{equation}
\label{eq:first_guess}
\hat{S}_{0}(a,b)=C(a,b)
\end{equation}

\begin{figure}[t!]
	\centering
	\includegraphics[height=5.5cm]{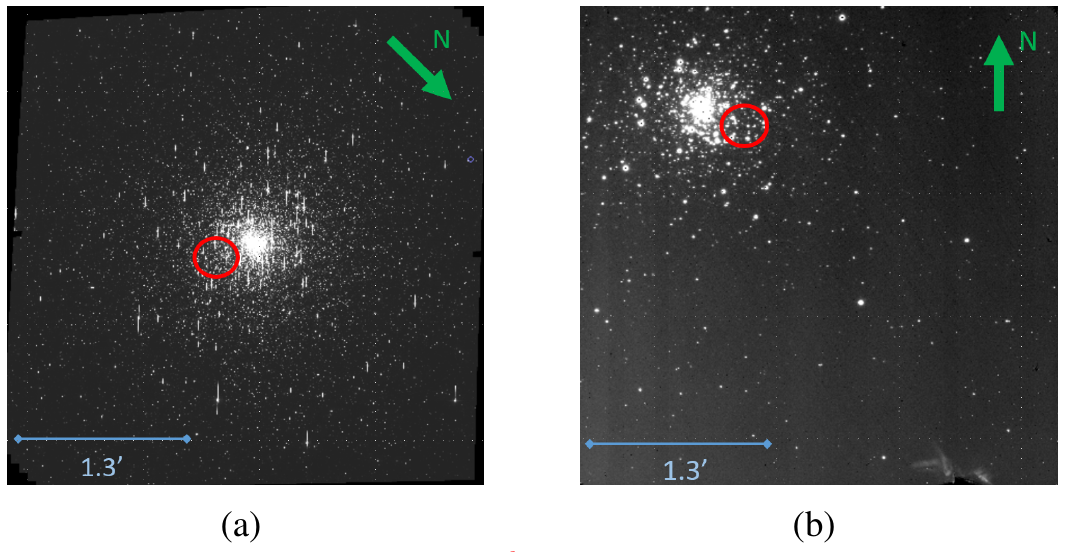}
	\caption{NGC1851 (Cladwell 73).  Located at $05^h 14^m 07^s ,-40^{\circ}02'48"$.  Red circle indicates region of interest where the binary is located. (a) MAST provided HST ACS WFC observation. (b) ESO provided HAWK-I 2.0-2.3 $\mu m$ observation}
	\label{fig:NGC1851}
\end{figure}

\begin{figure}[t!]
	\centering
	\includegraphics[height=5.5cm]{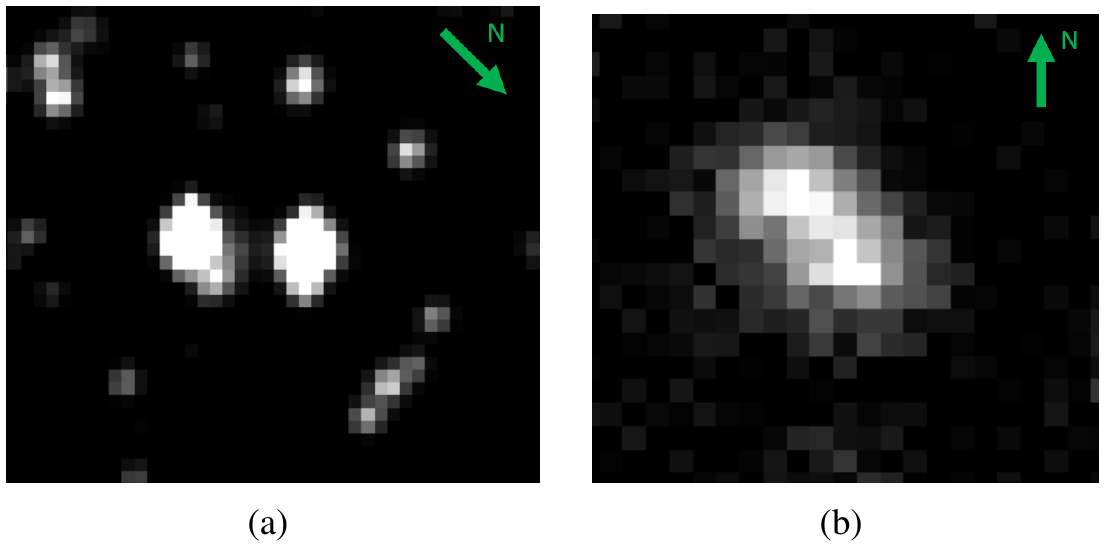}
	\caption{Binary of interest. (a) MAST provided HST ACS WFC observation. (b) ESO provided HAWK-I 2.0-2.3 $\mu m$ observation.}
	\label{fig:cropped}
\end{figure}

\begin{figure}[t!]
	\centering
	\includegraphics[height=4.5cm]{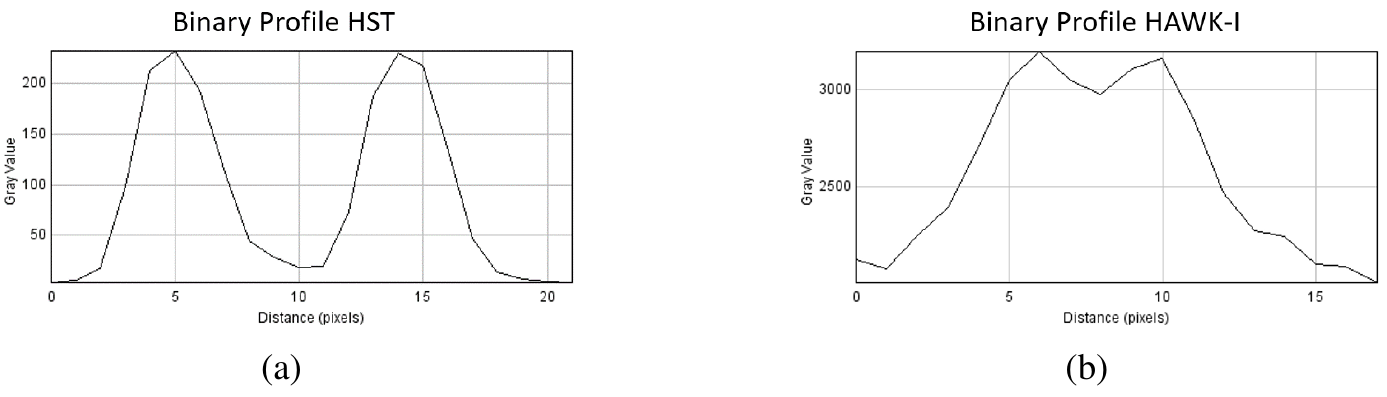}
	\caption{Profile across the binary in pixel space. Little confusion in HST image but significant region of overlap in HAWK-I image.  Images have distinct pixel scales. (a) HST (b) HAWK-I}
	\label{fig:cross_section}
\end{figure}

Here we will examine a resolvable binary source pair that has been observed by both the European Southern Observatory (ESO) instrument HAWK-I's hybridized H2RG HgCdTe 2.5$\mu$m array at the Very Large Telescope (VLT) and the CCD array of the Hubble Space Telescope's (HST) Advanced Camera for Surveys (ACS) Wide Field Camera (WFC).  The first binary explored here is located on the western side of NGC1851 with full frames shown in figure~\ref{fig:NGC1851} and the particular binary shown in figure~\ref{fig:cropped}.  It was selected for a number of reasons.  It is sufficiently above background in the HAWK-I exposure to be clearly visible.  The sources are visibly resolvable but extremely confused in the IR image as illustrated in figure~\ref{fig:cross_section}.  It was sufficiently below saturation in the HST image that minimal blooming occurs.

\begin{figure}
\begin{center}
\includegraphics[height=5.0cm]{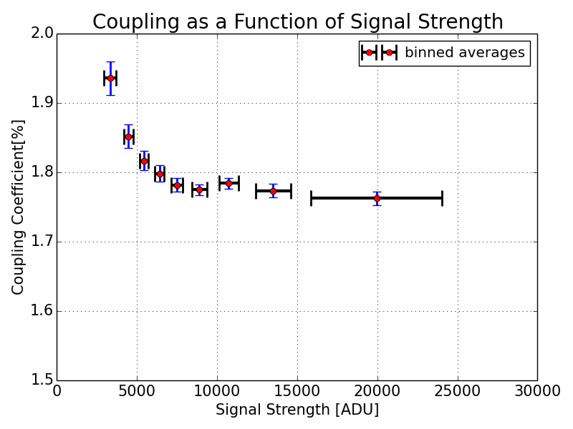}
\caption{Estimate of $\alpha(S)$ using hot pixels as isolated single pixel events above a dark background.  The form is the same as has been historically observed in this type of detector.  Coupling peaks on the order of 2.00\% tapering down to 1.75\%.  Vertical error bars indicate 98\% confidence on the means.  Horizontal error bars indicate $\pm$1$\sigma$ bounds on mean signal level binning.}
\label{fig:alph_hot}
\end{center}
\end{figure}

Point spread fitting using the python photutils implementation~\cite{Bradley16} of DAOphot~\cite{Stetson87} was performed to estimate the angular separation between the binary in the Hubble frame as well as the HAWK-I frame.  Theoretical predictions indicate that in arrays with IPC present, the separation between confused sources will be underestimated~\cite{Donlon18}.  This is due to the signal dependence of IPC causing preferential blurring towards the center of the binary compared to away.  IPC has been characterized as a function of signal strength using hot pixels present in dark exposures of the HAWK-I device as shown in figure~\ref{fig:alph_hot}.

This characterization will then be used to decouple the HAWK-I image.  Comparisons between the point spread fits from the original HAWK-I frame, the corrected HAWK-I frame, and the Hubble frame are shown in the following table.  The fit from the Hubble frame will be used as a type of fiducial, as there is no IPC present and astrometric errors will be considered as deviation from this data.

\begin{center}
\begin{tabular}{|c|c|c|c|}
\hline
 & HST & HAWK-I, uncorrected & HAWK-I, IPC corrected \\
\hline \hline
Separation [pix] & 9.265 & 4.288 & 4.320 \\
\hline
Separation [marcsec] & 463.251$\pm$0.583 & 456.363$\pm$3.149 & 459.823$\pm$3.149 \\
\hline
Error [marcsec] & --- & -6.888$\pm$3.203 & -3.428$\pm$3.203\\
\hline
Error [\%] & --- & -1.48\% & -0.74\% \\
\hline
\end{tabular}
 \end{center}


Separation of this confused binary using PSF fitting is measured to be smaller using data from a hybridized array compared to data from a CCD array. This underestimate is partially mitigated by correction using a partially characterized coupling coefficient.  It is worth noting that part of this precision comes from the underlying geometric distortions incident on the Hubble and HAWK arrays.  Raw HST images are anticipated to have up to $7\%$ geometric errors but these field distortions are corrected to less than $0.5 \ marcsecs$ by using the AstroDrizzle preprocessing technique \cite{Kozhurina18}.  On the VL 8m telescope the field correction is a more difficult problem as the IPC correction must be applied prior to any geometric correction.  As a result the raw uncorrected data was examined here.  This raw frame includes an expected residual error on the visual field on the order of $35 \ marcsec$. However, this isn't the error that we care about for separation fitting.  We need not be concerned with overall flatness of the field but only local flatness over a small neighborhood of $\sim 25 \ pixels$.  Examining the residuals presented in Libralto M, et al \cite{Libralato14}  indicates that astrometric accuracy across the field varies continiously with greater local change in residual error from distortions near the edge of the frame.  With a total astrometric error of $\pm250 \ marcsecs$ indicating an average per pixel change in geometric accuracy on the order of $0.0625 \ marcsecs$. These same residuals indicate that the peak distortion can be seen to be approximately double this average; a maximum per pixel change of $0.125 \ marcsecs$.  Extending that over the 25 pixel range, the maximum radius for confusion of pointspreads presented in this work, a maximum expected impact on separation of the geometric distortions on the order of $3.2 \ marcsecs$. The geometric distortion's contribution to separation error here is small.  This is prior to any geometric field correction, the application of a relatively simple 5th order polynomial type correction \cite{Libralato14} is a possibility but has not been used on the data presented here.  Using IPC correction that accounts for a change in coupling across the field reduces the magnitude of this underestimate from 1.48\% to 0.74\%.  This correction only used a signal dependent characterization of IPC from hot pixels;  it did not account for variation of IPC across the neighboring pixel level as well.  It is anticipated that a more complete characterization, such as that discussed later in this work, would allow for greater restoration of astrometric accuracy.

\subsection{Sets of binaries}
\begin{figure}[t!]
	\centering
	\includegraphics[height=5.5cm]{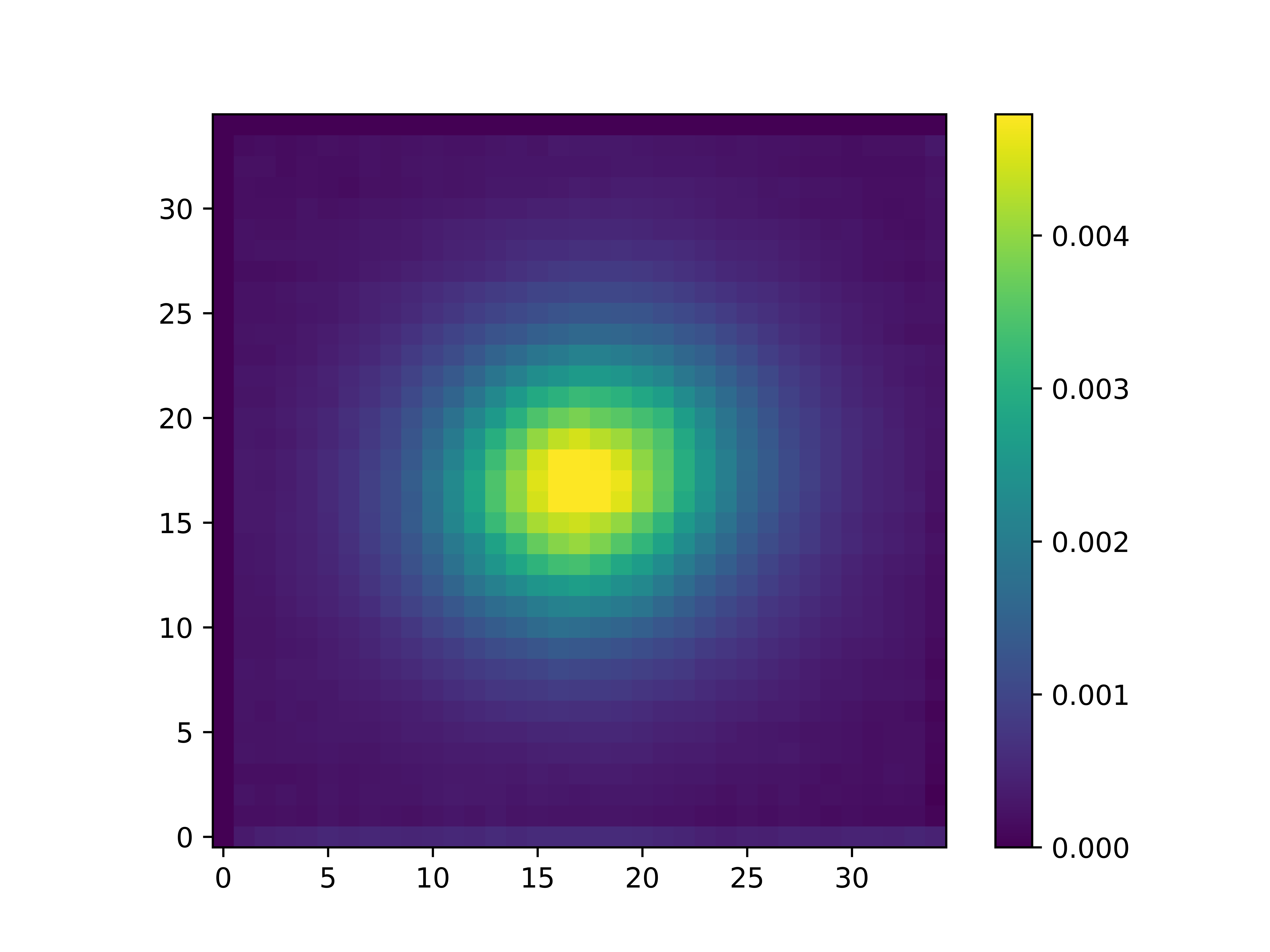}
	\caption{Constructred PSF from the NGC3603 frame for the HAWK-I array; raw and coupled are visually non-distinct.}
	\label{fig:HAWK_PSF_construction}
\end{figure}

\begin{figure}
	\begin{center}
	\includegraphics[height=5.0cm]{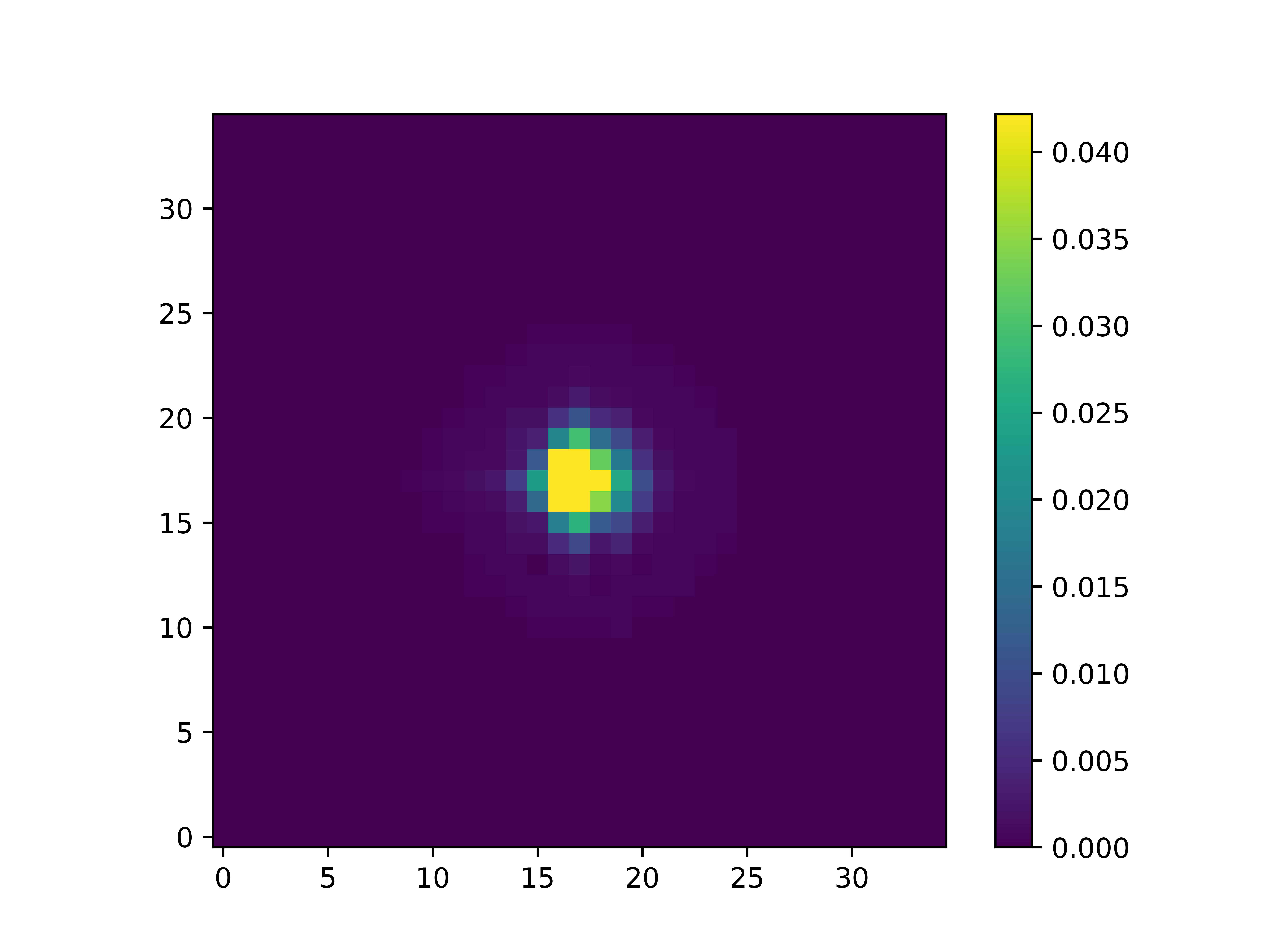}
	\caption{Constructed PSF from the NGC3603 frame for the HST array}
	\label{fig:HST_PSF_construction}
	\end{center}
\end{figure}

\begin{figure}[t!]
	\centering
	\includegraphics[height=5.5cm]{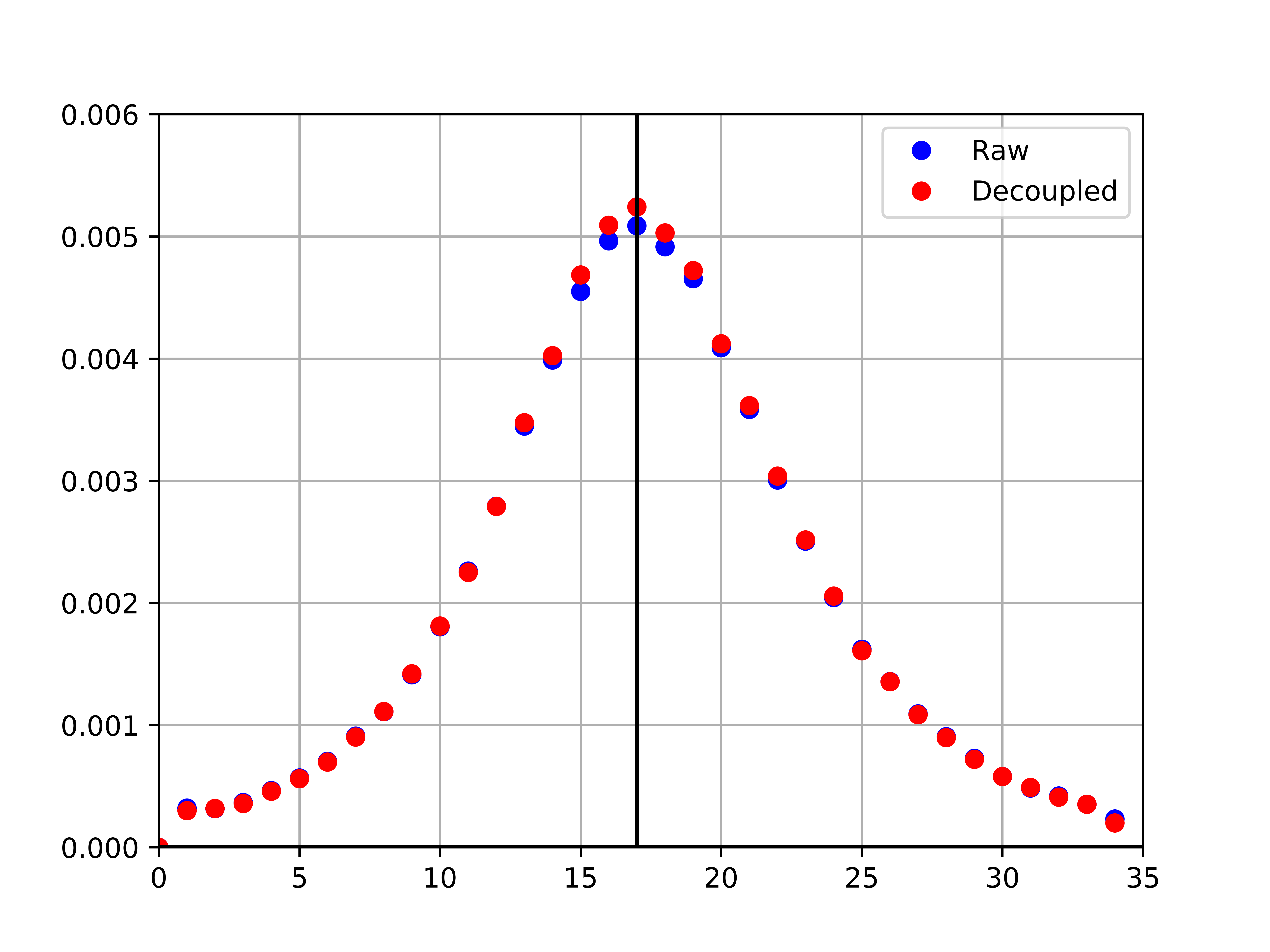}
	\caption{Constructed PSF cross section from the NGC3603 frame for the HAWK-I array.  Note the higher peak for the decoupled PSF compared to the raw.  Y-axis in units of fractional energy in pixel.  X-axis in units of pixels }
	\label{fig:HAWK_PSF_construction_cross}
\end{figure}

\begin{figure}
	\begin{center}
	\includegraphics[height=5.0cm]{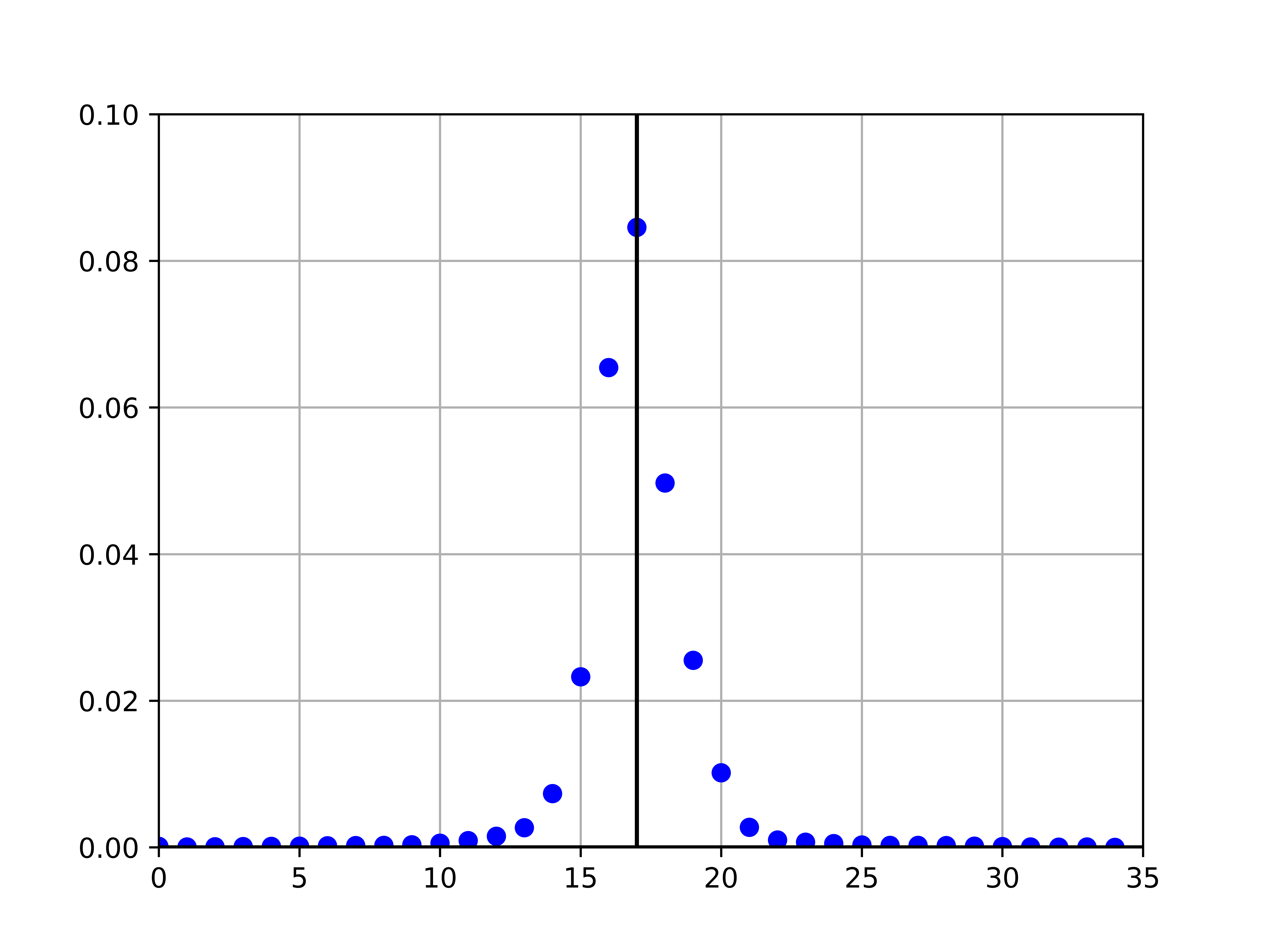}
	\caption{Constructed PSF cross section from the NGC3603 frame for the HST array. Y-axis in units of fractional energy in pixel.  X-axis in units of pixels}
	\label{fig:HST_PSF_construction_cross}
	\end{center}
\end{figure}

In addition to the NGC1851 data, another object, an open cluster NGC3603 located at $11^{h} \: 15^{m} \: 23^{s},$ $ -61^{\circ} 15' 00''$ was examined to increase the number of binaries.  The PSFs established from this frame are presented in figure \ref{fig:HAWK_PSF_construction} and \ref{fig:HST_PSF_construction}.  This same data is also illustrated as PSF cross sections in figures \ref{fig:HAWK_PSF_construction_cross} and \ref{fig:HST_PSF_construction_cross}.  This illustrates that the deconvolution impacts the measured PSFs in the anticipated way, by narrowing the FWHM and increasing energy in the peak.  The FHWM of the PSF from the NGC3603 frame is substantially larger ($\sim 11\: pixels \approx 1.166 \: arcsec$) than from the NGC1851 frame ($\sim 4\: pixels \approx 0.424\: arcsec$) due to  introduction of an adaptive optic system between the two observations.  The absence of this adaptive optic system is helpful when looking for confused sources, as a greater angular separation can give rise to the same degree of confusion.  The NGC3603 has resulted in 74 binary stars being fit.  These are characterized by the difference in measurement of angular separation between HAWK-I and HST.

\begin{figure}
	\begin{center}
	\includegraphics[height=8.0cm]{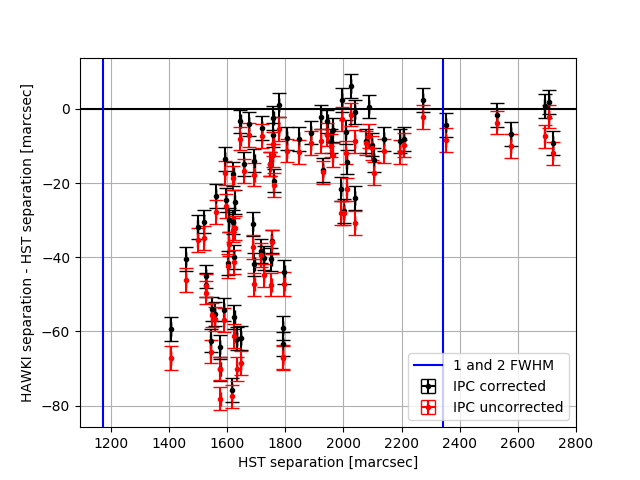}
	\caption{Error in separation from binary stars as a function of separation distance for corrected and uncorrected images of NGC3603.  74 stars are examined here.  The corrected data systematically increases the separation returned from PSF fitting.  In 72 of 74 cases, this resulted in a decrease in error relative to the uncorrected fit.  In the two excepting cases the deconvolution pushed the estimate of separation from a small underestimate to a small overestimate.}
	\label{fig:74_stars}
	\end{center}
\end{figure}

Figure \ref{fig:74_stars} illustrates a systematic decrease in separation error as a result of deconvolution of up to $8 \ marcsec$ with an average restoration on the order of $3.63 \ marcsec$.  This lack of complete correction is likely due to the lack of complete characterization of the coupling coefficient. More importantly, this figure shows that IPC will introduce a systematic error in astrometry on the order of tens of $marcsec$.  The greatest error occurs when the confusion is greatest and drops to zero when the sources are no-longer confused.  The data presented here is akin to and gives empirical validation to that simulated in previous work\cite{Donlon18} but does not have a clearly specified relationship between the relative intensities of the two sources or an absolutely defined peak intensity.  That is to say that this data is sampling from higher dimensional curves in both absolute and relative brightness.  The dropoff in separation error as confusion of the sources decreases is clear; separation trends to nearly no astrometric error compared to the hubble data when the sources are distinct.

\section{Characterization using single pixel resets}
The technique applied in section one of this paper is preferable serves to mitigate the impact of IPC coupling on astrometric error but does not eliminate it completely.  This is largely due to the assumption inherent to using hot pixel characterization. Hot pixels only provide a cross section through equation ~\ref{eq:algebra_solved} where $\alpha(S|B=0)$ rather than the full $\alpha(S,B)$.  What follows outlines a method of using single pixel resets to characterize IPC more accutately over a more representitive domain.  This higher quality characterization feeds into a more accurate correction, yielding further minimization of astrometric errors.

\begin{figure}[t!]
	\centering
	\includegraphics[height=4.5cm]{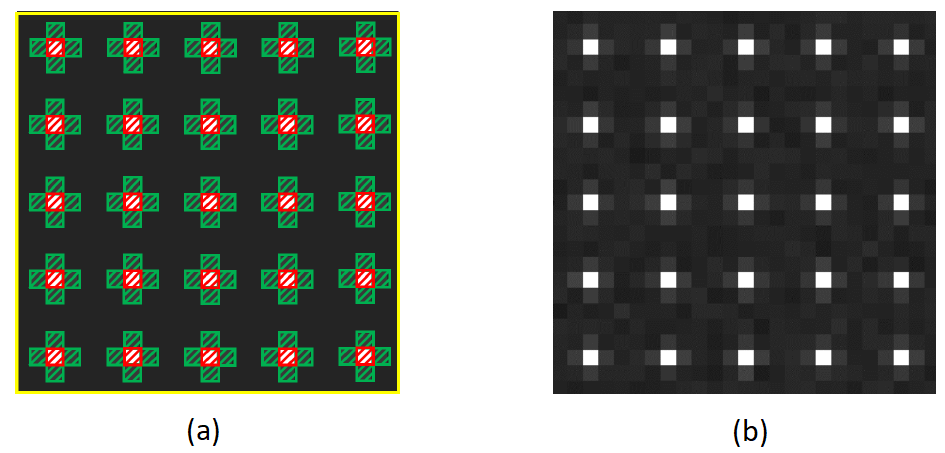}
	\caption{(a) An illustration of SPR samplings in the absence of noise demonstrating the four sorts of pixel contained.  Red pixels are instances of $C$, green are instances of $N$, all others are instances of $B$. (b) sample SPR frame with IPC, shot noise on the background, and read noise.}
	\label{fig:SPRs}
\end{figure}

\label{sect:SPR}
	In order to characterize the coupling coefficient, single pixel events can be examined \cite{Seshadri08}. Hot pixels over dark frames are an effective technique to characterize IPC coupling as a function of signal level but cannot explore the dependence on the depletion state of the neighboring pixels\cite{Donlon16} , as in order to be an isolated event in a dark frame, the neighbor level must be approximately zero.  A proposed alternative method is to use single pixel resets.  These hybridized arrays allow for pixels to be addressed and reset individually.  Extending the technique of SPR to a full characterization can be done through the following method:
\begin{enumerate}
	\item Reset the array to prepare for an exposure.
	\item Expose the array to a flat field for to produce a particular background level.
	\item Reset isolated pixels using a voltage $V_{sig}$ corresponding to a particular level of depletion here called 'signal'.
	\item Read out the array and repeat for new $V_{sig}$ and/or exposure time.
\end{enumerate}
A sample for this type of data can be seen in figure~\ref{fig:SPRs}.  Considering only nearest neighbor coupling, each frame obtained using this method where $n$ isolated pixels are reset with a center-to-center gridded separation of $m$ pixels in a square grid you acquire n samples of your $C$ value, $4n$ samples of your $N$ value, and  $(m^{2}-5)n$ samples of your $B$ value.  These sample counts are what will limit statistical confidence on results.  Instances of each of these data types are illustrated in figure~\ref{fig:SPRs}b.

	  From this type of frame a scatter of points can be built up using:
\begin{equation}
\label{eq:observations}
\langle \alpha \rangle = \frac{ \langle N \rangle - MED(B)}{\langle C \rangle +4 \langle N \rangle - 5(MED(B))}
\end{equation}
Where $MED(B)$ is the median of the set of $B$ samples and $\langle X \rangle$ indicates the mean of the set of samples of $X$ and with expected noise statistics governed by~\cite{Ku66}:
\begin{equation}
\label{eq:Jac}
\sigma ^2 _{\alpha} \approx \vec{J}\mathbf{\Sigma}^{\vec{x}}\vec{J}^T
\end{equation}
\tiny
\begin{equation}
\label{eq:noise_expanded}
\sigma^2_{\alpha} \approx \frac{( \langle C \rangle - \langle B \rangle ) ^2 \sigma ^2 _N + ( \langle N \rangle - \langle C \rangle ) ^2 \sigma ^2 _B + ( \langle N \rangle - \langle B \rangle ) ^2 \sigma ^2 _C + 2 ( \langle C \rangle - \langle B \rangle ) ( \langle N \rangle \ - \langle C \rangle ) \sigma _{N,B} + 2 ( \langle C \rangle - \langle B \rangle ) ( \langle N \rangle \ - \langle B \rangle ) \sigma _{N,C} + 2 ( \langle N \rangle - \langle C \rangle ) ( \langle N \rangle \ - \langle B \rangle ) \sigma _{B,C}}{( \langle C \rangle + 4 \langle N \rangle - 5 \langle B \rangle ) ^{4}}
\end{equation}
\normalsize
	Sensibly, this noise blows up when $C \approx B$, as the read out does not indicate a  uniquely defined coupling coefficient in this circumstance.  In order to characterize the coupling coefficient in this neighborhood, interpolation must be used.  To examine the behavior of this technique,  it has been applied to simulated data with a known coupling coefficient prescribed and applied. The form used for this simulated coupling coefficient was as follows:
\begin{equation}
\label{eq:alpha_S_B}  
\alpha ( S , B ) = A_0 exp\left( -\frac{|S-B|}{k_0} \right) + A_1 exp \left( - \frac{(S^2 + B^2)^{0.5}}{k_1} \right) + \alpha_{\infty}
\end{equation}
	This form is approximate but is informed by observations~\cite{Cheng09} and simulations~\cite{Donlon17} as well as constraining that the coupling coefficient from pixel $i$ to $j$ is identical to the coupling coefficient from pixel $j$ to $i$.  

These simulated frames were created using the following method:
\begin{enumerate}
	\item Generate a uniform background level.
	\item Apply shot noise to each sample of background level.
	\item Set fixed pixels to a reset level.
	\item Simulate IPC coupling through application of equation~\ref{eq:conv} with coupling defined by equation~\ref{eq:alpha_S_B}
	\item Apply zero-mean normal read noise to each sample.
	\item Repeat for many background and reset levels.
\end{enumerate}

\begin{figure}[t!]
	\centering
	\includegraphics[height=4.5cm]{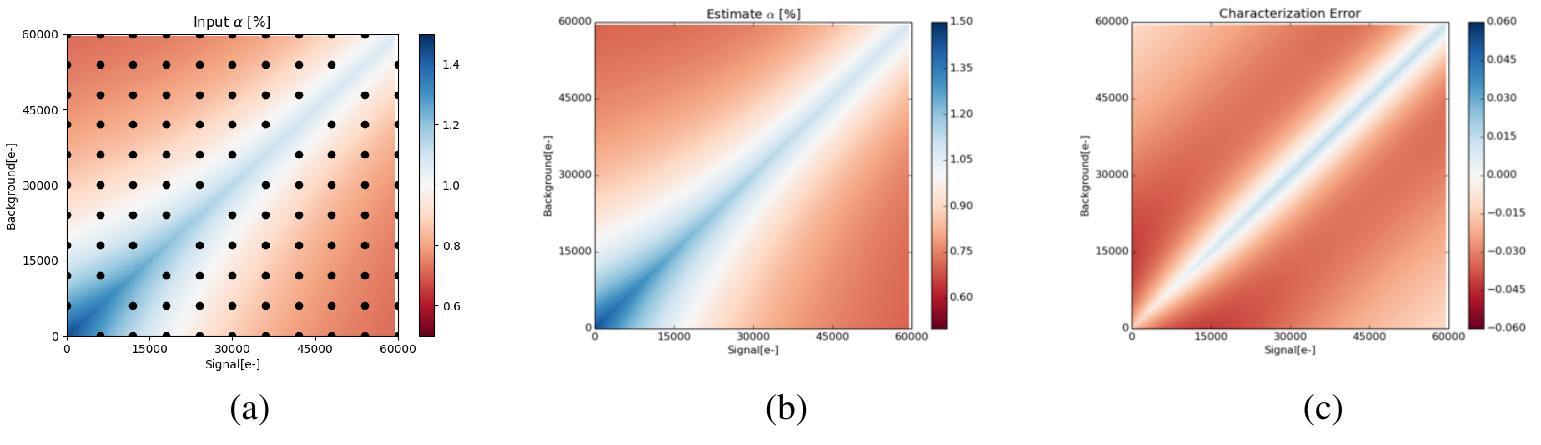}
	\caption{(a) Input $\alpha (S,B)$ as prescribed by equation~\ref{eq:alpha_S_B} with sampled frames collected at the illustrated points.  Note that no frames were examined where $S=B$ due to the non-uniqueness of equation~\ref{eq:alpha_S_B} at those points.   (b) Regressed form of $\alpha (S,B)$ using data frames described to fit to equation~\ref{eq:alpha_S_B}. (c) Difference between (b) and (a) $Regressed-Input$.  Peak overestimate in $\alpha$ of 0.020[\%] and peak underestimate of 0.045[\%].}
	\label{fig:projection}
\end{figure}

This technique preserves the property of IPC that signal and shot noise are coupled, but read noise is not.  Every reset, through equation~\ref{eq:observations} gives rise to an observation of $(\alpha| S, B)$.  Assembling all this data yields a scatter plot in three dimensions. A simulated dataset was examined by processing 100 frames at each point noted in \ref{fig:projection} a with each frame containing 9,604 resets with 30RMS simulated read noise.  Fitting this data to the prescribed form yields coefficients and errors summarized in the following table.

\begin{center}
\begin{tabular}{|c|c|c|c|}
\hline
Parameter & Input & Estimate & Fractional Error \\
\hline \hline
$A_0$ & 0.400[\%] & 0.416[\%] & +0.040 \\
\hline
$A_1$ & 0.400[\%] & 0.375[\%] & -0.063 \\
\hline
$\alpha_\infty$ & 0.650[\%] & 0.644[\%] & -0.009 \\
\hline
$k_0$ & 20,000 [e-] & 15,573 [e-] & -0.221 \\
\hline
$k_1$ & 28,284 [e-] & 30,816 [e-] & -0.090 \\
\hline
\end{tabular}
 \end{center}
Though the error on any individual parameter can be quite large, the errors compensate for each other yielding a maximum overestimate in fractional IPC of 0.02[\%] and maximum underestimate of 0.045[\%] for coupling scaling from 0.65[\%] to 1.45[\%].

	The behavior of this fit is illustrated in figure~\ref{fig:projection}b with figure~\ref{fig:projection}c showing the difference between the input coupling and the regressed coupling as a function of $S$ and $B$.

\section{Summary}
	A particular set of binaries is examined where the impact of IPC on separation estimates is indicated to be significant.  PSF fitting techniques corroborate this claim when compared to fiducial data from HST.  Correction of IPC using a partial hot pixel based characterization of the coupling coefficient is performed.  The corrected data exhibits a smaller underestimate of separation indicating incomplete restoration of astrometric accuracy. A technique by which IPC can be characterized as a function of pixel and background level using single pixel resets is explored.  The noise behavior and statistical properties of this type of characterization are explored using simulated data.

	When examining dense fields, where sources are not well isolated, PSF fitting is the dominant technique to obtain astrometric and photometric measurements.  IPC causes distortions to the PSF as the signal integrated on a pixel changes;  bright sources appear more narrow than faint sources.  Furthermore, when the sources are significantly confused, the PSFs distort asymmetrically.  This work shows that these distortions can cause inaccuracy of astrometric measurements as well as demonstrates partial success of a correction algorithm when using a partial characterization of IPC.  Future work will explore this error more systematically on sets of binaries to establish trends and better inform astronomers of the degree to which IPC will impact their measurements for particular observations.

\acknowledgments 
	We would like to thank NASA for providing support and funding through contract NAS5-02105.  Some of the data presented in this paper were obtained from the Mikulski Archive for Space Telescopes (MAST).  STScI is operated by the Association of Universities for Research in Astronomy, Inc., under NASA contract NAS5-26555.  Based on observations made with the NASA/ESA Hubble Space Telescope, and obtained from the Hubble Legacy Archive, which is a collaboration between the Space Telescope Science Institute (STScI/NASA), the European Space Agency (ST-ECF/ESAC/ESA) and the Canadian Astronomy Data Centre (CADC/NRC/CSA).  Based on data obtained from the ESO Science Archive Facility under request numbers Donlok 321673 and Donlok 321674.

\bibliography{./report.bib} 

\begin{thebibliography}{10}

\bibitem{Rogalski00}
A.~Rogalski, K.~Adamiec, and J.~Rutkowski, {\em Narrow-Gap Semiconductor
  Photodiodes}, SPIE Publications, Bellingham Washington  (2000).

\bibitem{Moore06}
A.~C. Moore, Z.~Ninkov, and W.~J. Forrest, ``Quantum efficiency overestimation
  and deterministic cross talk resulting from interpixel capacitance,'' {\em
  Opt. Eng.} {\bf 45}, 076402  (2006).

\bibitem{Ohanian07}
H.~C. Ohanian, {\em Classical Electrodynamics}, Infinity Science Press, Hingham
  Massachusetts, 2~ed.  (2007).

\bibitem{Donlon17}
K.~Donlon, Z.~Ninkov, S.~Baum, and L.~Cheng, ``Modeling of hybridized infrared
  arrays for characterization of interpixel capacitive coupling,'' {\em Optical
  Engineering} {\bf 56}  (2017).

\bibitem{Cheng09}
L.~Cheng, ``Interpixel capacitive coupling,'' Master's thesis, Rochester
  Institute of Technology  (2009).

\bibitem{Donlon16}
K.~Donlon, Z.~Ninkov, and S.~Baum, ``Signal dependence of inter-pixel
  capacitance in hybridized {HgCdTe} {H2RG} arrays for use in {James Webb}
  space telescope's {NIRcam},'' in {\em High Energy Optical and Infrared
  Detectors for Astronomy VII},  A.~D. Holland and J.~Beletic, Eds., {\em Proc.
  SPIE} {\bf 9915}  (2016).

\bibitem{Moore04}
A.~C. Moore, Z.~Ninkov, and W.~J. Forrest, ``Interpixel capacitance in
  non-destructive focal plane arrays,'' in {\em Focal Plane Arrays for Space
  Telescopes},  T.~J. Grycewicz and C.~R. McCreight, Eds., {\em Proc. SPIE}
  {\bf 5167}, 204--215  (2004).

\bibitem{Seshadri08}
S.~Seshadri, D.~M. Cole, B.~R. Hancock, and R.~M. Smith, ``Mapping electrical
  crosstalk in pixelated sensor arrays,'' in {\em High Energy, Optical, and
  Infrared Detectors for Astronomy III},  {\em Proc. SPIE} {\bf 7021}  (2008).

\bibitem{Ku66}
H.~H. Ku, ``Notes on the use of propagation of error formulas,'' {\em National
  Bureau of Standards} {\bf 70C}, 263  (1966).

\bibitem{Donlon18}
K.~Donlon, Z.~Ninkov, and S.~Baum, ``Point-spread function ramifications and
  deconvolution of a signal dependent blur kernel due to interpixel capacitive
  coupling,'' {\em Publications of the Astronomical Society of the Pacific}
  {\bf 130}  (2018).

\bibitem{Bradley16}
L.~Bradley, B.~Sipocz, T.~Robitaille, and E.~T. et~al.,
  ``astropy/photutils:v0.3,''  (2016).

\bibitem{Stetson87}
P.~B. Stetson, ``{DAOPHOT} - a computer program for crowded-field stellar
  photometry,'' {\em Publications of the Astronomical Society of the Pacific}
  {\bf 99}, 191--222  (1987).

\bibitem{Libralato14}
M.~Libralato, A.~Bellini, L.~R. Bedin, G.~Piotto, I.~Platais, M.~Kissler-Patig,
  and A.~P. Milone, ``Ground-based astrometry with wide field imagers v.
  application to near-infrared detectors:{HAWK-I@VLT/ESO},'' {\em Astronomy and
  Astrophysics} {\bf 563}  (2014).

\bibitem{Kozhurina18}
V.~Kozhurina-Platais, N.~Grogin, and E.~Sabbi, ``Accuracy of the {HST} standard
  astrometric catalogs w.r.t. gaia,'' Tech. Rep. ACS/WFC 2018-01, Space
  Telescope Science Institute  (2018).

\bibitem{Donlon18_2}
K.~Donlon, Z.~Ninkov, and S.~Baum, ``Signal dependent interpixel capacitance in
  hybridized arrays: simulation, characterization, and correction,'' in {\em
  High Energy Optical and Infrared Detectors for Astronomy VIII},  A.~D.
  Holland and J.~Beletic, Eds., {\em Proc. SPIE} {\bf 10709}  (2018).

\end{thebibliography}
\bibliographystyle{./spiejour.bst} 


\end{spacing}

\end{document}